\documentclass[pdftex,citeautoscript,aps,pra,superscriptaddress,cite,amsmath,amsfonts,amssymb,twocolumn]{revtex4}
\pdfoutput=1
\renewcommand{\vec}[1]{\mathbf{#1}} 
\usepackage{graphicx}
\usepackage{color}
\usepackage{url}

\newcommand{\figref}[1]{Fig.~\ref{fig:#1}}

\newcommand{\eqnumref}[1]{(\ref{eq:#1})}
\renewcommand{\eqref}[1]{Eq.~\eqnumref{#1}}

\newcommand{\citeasnoun}[1]{Ref.~\onlinecite{#1}}

\newcommand{\secref}[1]{Sec.~\ref{sec:#1}}

\begin{document}
\title{Casimir repulsion beyond the dipole regime}
\author{Alexander~P.~McCauley}
\affiliation{Department of Physics, Massachusetts Institute of Technology, Cambridge, MA 02139, USA}
\author{Alejandro~W.~Rodriguez}
\affiliation{School of Engineering and Applied Sciences, Harvard University, Cambridge, MA 02139, USA}
\affiliation{Department of Mathematics, Massachusetts Institute of Technology, Cambridge, MA 02139, USA}
\author{M.~T.~Homer~Reid}
\affiliation{Department of Physics, Massachusetts Institute of Technology, Cambridge, MA 02139, USA}
\affiliation{Research Laboratory of Electronics, Massachusetts Institute of Technology, Cambridge, MA 02139, USA}
\author{Steven~G.~Johnson}
\affiliation{Department of Mathematics, Massachusetts Institute of Technology, Cambridge, MA 02139, USA}


\begin{abstract}
We extend a previous result [Phys. Rev. Lett. 105, 090403 (2010)] on
Casimir repulsion between a plate with a hole and a cylinder centered
above it to geometries in which the central object can no longer be
treated as a point dipole.  We show through numerical calculations
that as the distance between the plate and central object decreases,
there is an intermediate regime in which the repulsive force increases
dramatically.  Beyond this, the force rapidly switches over to
attraction as the separation decreases further to zero, in line with
the proximity force approximation.  We demonstrate that this effect
can be understood as a competition between an increased repulsion due
to a larger polarizability of the central object interacting with
increased fringing fields near the edge of the plate, and attractive
forces due primarily to the nonzero thickness of the plate.  In
comparison with our previous work, we find that using the same plate
geometry but replacing the single cylinder with a ring of cylinders,
or more generally an extended uniaxial conductor, the repulsive force
can be enhanced by a factor of approximately $10^3$.  We conclude that
this enhancement, although quite dramatic, is still too small to yield
detectable repulsive Casimir forces.
\end{abstract}
\maketitle

\section{Introduction}

\begin{figure}[tb]
\includegraphics[width=1.0\columnwidth]{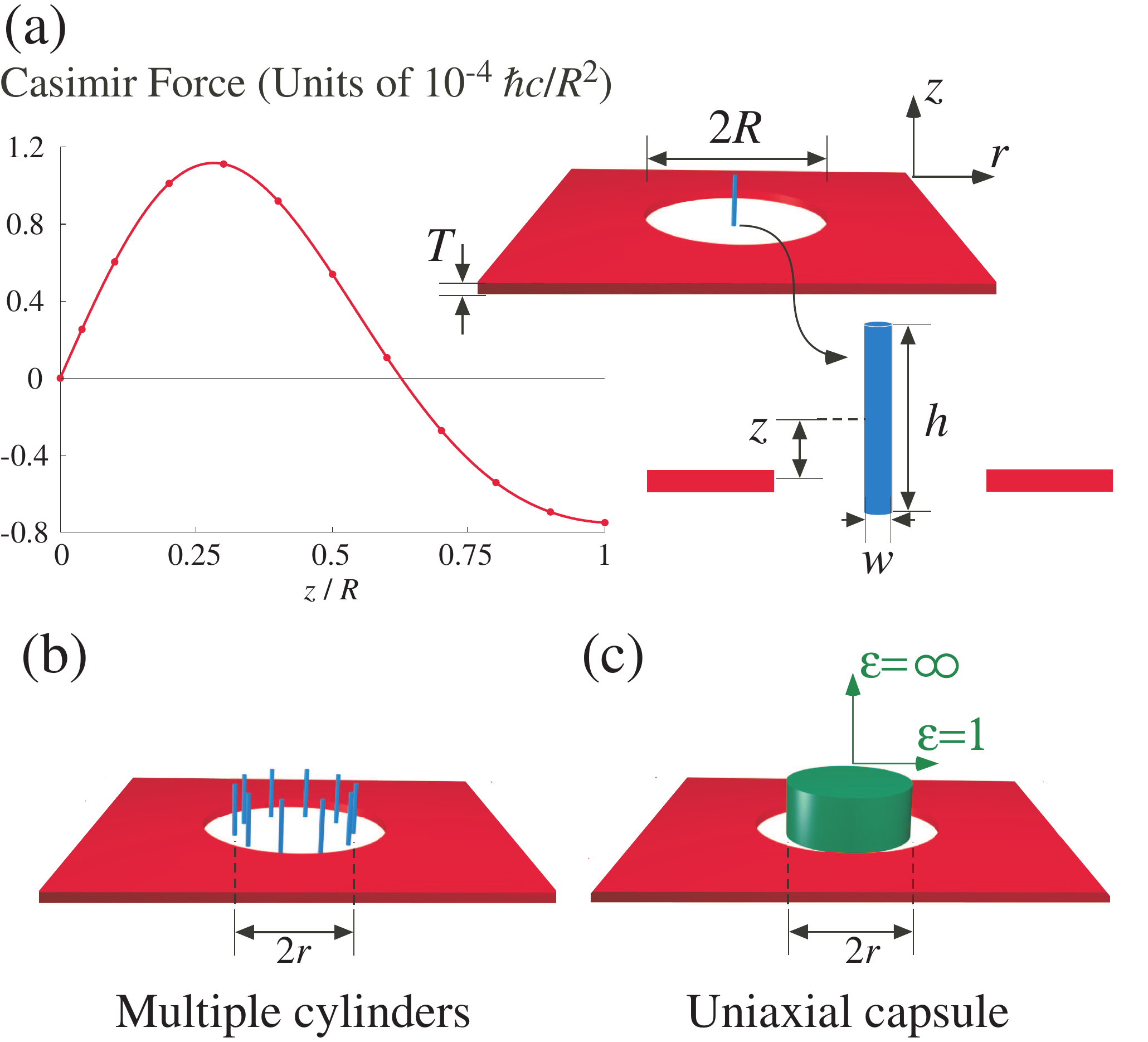}
\centering
\caption{Configurations used in the text.  \emph{Top}: Reference
  configuration, from~\citeasnoun{LevinMc10}.  \emph{Top Left}:
  Geometry illustrating the single plate-needle configuration.
  \emph{Top Right}: Casimir force vs. vertical height $z$ for the
  configuration in the top left panel.  Positive forces denote
  repulsion.  \emph{Bottom}: The two generalizations considered here:
  \emph{Bottom Left}: Replacing a single cylinder with a ring of $N$
  (in the figure, $N=10$) cylinders, each of which is more closely
  spaced to the edge of the hole.  \emph{Bottom Right}: An idealized
  version of the cylinder ring configuration, consisting of a cylinder
  of uniaxial conductance in the vertical ($z$) direction.}
\label{fig:configs}
\end{figure}

Although Casimir forces (a generalization of van der Waals forces to
macroscopic objects) between neutral metal objects in vacuum are
normally attractive interactions~\cite{milton01, RodriguezCa11}, in
our previous work we showed that the force can become repulsive for
objects of certain shapes, and in particular we showed that repulsion
occurred for a needle-like particle above a metal plate with a
hole~[\figref{configs}~(a)] for which an analytical symmetry argument
applied~\cite{LevinMc10}.  In this work, we address two questions:
first, is repulsion limited to systems where one particle is very
small or can it be obtained for two objects of length scales
comparable to their separation; and second, is this repulsion
necessarily weak ($10\,\mathrm{aN}$ for a single particle
in~\citeasnoun{LevinMc10}) or can it theoretically be made stronger in
comparison with other Casimir forces without simply shrinking the
entire system (for perfect conductors, multiplying all dimensions by a
scale factor $a$ changes the force by a factor of $1/a^2$, but in
actual systems there is a limit to the minimum achievable length
scales).  In answer to these questions, we find that a thousand-fold
enhancement of the repulsion is theoretically possible without
changing the hole radius or overall length scales.  This is
accomplished by replacing the original needle-like particle
of~\figref{configs}(a) (in this case a circular cylinder of high
aspect ratio) with a macroscopic array of identical
particles~[\figref{configs}(b)], effectively forming a capsule of
uniaxial conducting material~[\figref{configs}(c)].  To understand the
mechanism behind this enhancement, we separately consider the
repulsion as a single cylinder is displaced off-axis as well as the
screening interaction when multiple cylinders are combined.  For a
single cylinder we find that there is an optimal off-axis position
where the repulsion is enhanced (up to 50 times) due to the presence
of strong fringing fields.  The presence of this optimal position is a
consequence of the competing repulsive and attractive effects of
fringing fields and proximity-force approximation
(PFA)~\cite{Derjaguin34} interactions, respectively.  As the cylinder
is brought closer to the plate, the PFA forces take over and the
repulsion vanishes.  The Casimir force is non-additive, and the total
force between the array of cylinders and the plate will not simply be
a sum of the forces between the individual cylinders and the plate.
However, we find that the non-additive contributions (or screening
effects) associated with the presence of multiple cylinders are
surprisingly small: the enhancement does not saturate until there are
dozens of cylinders with separation significantly smaller than the
cylinder length.  We furthermore find that this repulsion persists
when realistic conductivities are included, and when the capsule is
anchored to a substrate by via a long oxide post.  Although
substantial experimental challenges remain in fabricating this
particular structure, these results, combined with previous work on
effective uniaxial/anisotropic media~\cite{Enk95:torque, Shao05,
  Munday05, Kenneth00:torque, Rodrigues06:torque,
  McCauley11:orientation} suggests that strongly anisotropic
metamaterials offer new opportunities to achieve exotic Casimir
interactions.

To begin with, we review the argument of~\citeasnoun{LevinMc10} that
describes repulsion in the dipole regime.  Consider the setup
of~\figref{configs}~(a), consisting of a thin plate with a hole of
radius $R$ and a thin cylinder with center a distance $z$ above the
hole and oriented normal to the plate.  Assume that the cylinder is
exactly centered on the hole, so that by cylindrical symmetry the
total force is parallel to the $z$ axis.  In the limit that this
cylinder is infinitesimal, only dipolar charge/current fluctuations on
the cylinder are allowed, and the energy is given by the
Casimir-Polder energy of a point dipole across from the hole.  If the
plate is perfectly conducting but has zero thickness
$T=0$~\endnote{Formally, we must take the perfectly conducting limit
  before the $T\rightarrow 0$ limit}, a repulsive force follows from a
symmetry argument: when the needle is at $z=0$, dipolar symmetry
prevents charge fluctuations of the needle from coupling to the plate
and vice versa.  The Casimir-Polder energy at $z=0$ must therefore be
zero. As $z\rightarrow \infty$ (fixing $\vec{x}=0$), the
(Casimir-Polder) energy must be negative, as in this limit the
geometry is equivalent to a needle and a uniform plate.  Therefore, at
some point the derivative of the energy with respect to $z$ must be
negative, implying a net repulsive force.  This argument is rigorously
true for an infinitesimal needle centered at $\vec{x}=0$ across from a
perfectly thin plate.  When the needle is of finite size $h>0$, but
still of a high aspect ratio $h \gg w$ (e.g., a long cylinder, so as
to be primarily polarizable in the $z$-axis), and the plate is of
finite thickness but with $R \gg T$, we found
in~\citeasnoun{LevinMc10} that the repulsive force persists as long as
$h \lesssim R/2$.

This repulsive force is interesting in that it can neither be
interpreted as arising (qualitatively) from pairwise-additive
forces~\cite{emig03_1, Emig07:ratchet, Neto08, Lambrecht09, Chiu10} or
from an effective-medium interpretation~\cite{Munday05, Leonhardt07,
  Romanowsky08, Rosa09, Zhao09} (for other such exceptions,
see~\citeasnoun{RodriguezCa11}).  However, as found
in~\citeasnoun{LevinMc10}, the magnitude of this repulsive force is
extremely small (on the order of $10\,\mathrm{aN}$ for the geometry
considered in that work), due to both the smallness of the needle and
its large distance ($R \sim 500\,\mathrm{nm}$) from the edge of the
plate.  Therefore, aside from examining the repulsive effect beyond
the dipole regime, another motivation of this work is to examine how
greatly the repulsive force can be enhanced.  An obvious way to
increase the absolute magnitude of the force would be to shrink all
dimensions of the system uniformly by a scaling factor $a$---for
perfect conductors, the force will scale as $1/a^2$ by dimensional
analysis.  However, for real materials there is a lower limit to the
dimensions that can be achieved before material effects (e.g., the
skin depth for gold) become important.  For the present case, the most
important geometric parameter is the thickness $T$ of the plate, which
must be nonzero for any physical configuration---we expect that $T$
should be larger than the skin depth of gold in order for our analysis
to be valid.  As discussed in~\citeasnoun{LevinMc10}, the repulsive
effect requires $T/R \ll 1$; setting, e.g., $T\sim 20\,\mathrm{nm}$ as
in~\citeasnoun{LevinMc10} as an optimistic lower bound on a real plate
thickness constrains the minimal value for $R$, and hence the overall
force magnitude.

The basic principle of repulsion in this system lies in the charge
fluctuations of the plane decoupling from fluctuations of the needle
at $z = 0$.  Then when considering more general geometries we should
have in mind a cylindrically-symmetric object of conductivity
primarily along the $z$-axis, which is our motivation for considering
the configurations of~\figref{configs}~(c,d).  In both cases, due to
the close contact between the center object and the edge of the plate,
it is essential to account for finite-size effects, making analytic or
semi-analytic calculation difficult.  Furthermore, in the presence of
a thick plate we cannot rely on a general symmetry argument to
guarantee repulsion for any range of parameters.  As a result, the
results of this work are entirely numerical.  The two computational
tools we use for our analysis are a boundary-element method
(BEM)~\cite{ReidRo09} and a finite-difference time-domain (FDTD)
method~\cite{RodriguezMc09:PRA, McCauleyRo10:PRA}, both of which
involve the adaptation of numerical techniques from classical
electromagnetism for Casimir force computation.  For simplicity, we
primarily assume perfectly-conducting materials, and we further drop
the requirement of a separating plane between the two objects (while
still requiring that the cylindrical object be more than halfway out
of the plane, so that the ``repulsion'' is not due to a trivial
redefinition of coordinates).  Additionally, to make a fair definition
of ``enhancement'', we work within geometric constraints for the plate
similar to~\citeasnoun{LevinMc10}.  One last fact bears mentioning
before proceeding further: in the cylindrically symmetric case, the
net Casimir force is along the $z$-axis.  However, this force is
intrinsically unstable~\cite{Rahi10:PRL}: a small radial displacement
of any of the configurations of~\figref{configs} off the $z$-axis will
result in a large, attractive radial component to the Casimir force.
Therefore, to prevent this translational instability any configuration
must be confined in the radial direction by, e.g., external mechanical
forces.

\section{Enhancement of fringing fields}

\begin{figure}[tb]
\includegraphics[width=1.0\columnwidth]{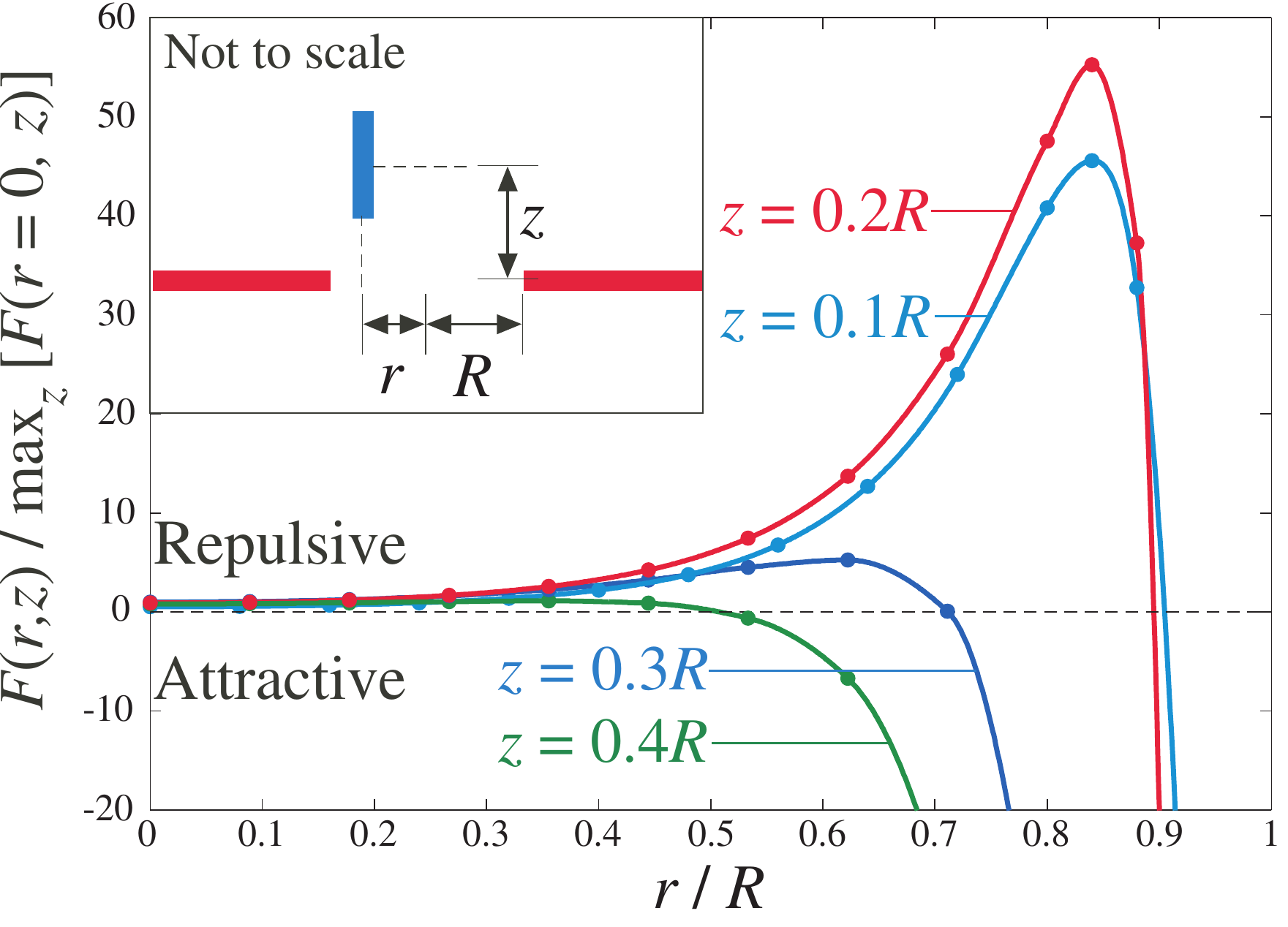}
\centering
\caption{The effect of fringing fields on $F_z$, the $z$-component of
  the Casimir force for a single metallic cylinder: for a fixed $z$,
  $F_z$ is enhanced as $r$ increases.  Plotted above is $F_z$ for
  several values of $z/R$, normalized by the maximum of the repulsive
  force ($0.0278\, \hbar c / R^2$) for $x = 0$.  As $r\rightarrow R$,
  $F_z$ undergoes a very strong enhancement, peaking at approximately
  55 for $r\sim 0.8R$ and $z = 0.2R$.  For every $z$, as $x$ increases
  pairwise attractive forces take over and $F_z$ switches sign (note
  that for $r\neq 0$ there is an attractive force $F_x$ between the
  cylinder and the plate edge, so the total Casimir force is not
  strictly repulsive in this case).  Shown for reference in the inset
  is $F_z$ (in units of $\hbar c/R^2$) at $r=0$.}
\label{fig:single-cylinder}
\end{figure}

As noted in~\cite{LevinMc10}, the repulsive forces arise from fringing
fields due to the sharp edges around the hole of the plate; this holds
for both perfectly thin ($T=0$) and thicker plates.  Drawing on our
intuition from electrostatics, for any fixed plate geometry we suspect
that these fields are greatly enhanced as the plate edges are
approached (i.e., $r = |\vec{r}| > 0$).  If the plate thickness is
\emph{a priori} fixed, it is possible that this enhancement of the
fringing fields leads to an enhanced repulsive effect for some range
of $r$.  Because the repulsive force is unstable in the transverse
direction, a single cylinder displaced by $r>0$ will experience a
strong (attractive) radial force; to compensate for this, we would
need to include, e.g., multiple cylinders displaced symmetrically
about $r = 0$ to make the net radial force zero.  However, to
understand the effects of the fringing fields alone it is more
illuminating to first consider the $z$-component $F_z(r,z)$ of the
Casimir force, as a function of $r$ and $z$, between a \emph{single}
cylinder at position $(r,z)$ and a \emph{thin} ($T=0$) plate, as
depicted in~\figref{single-cylinder} (Inset).  For definiteness, the
cylinder is taken to have height $h = 0.64 R$ and width $w = 0.04 R$
(the same units as in~\citeasnoun{LevinMc10}).  In this case, the
plate is taken to be a circular annulus of inner radius $R$ and outer
radius $8R$, the latter being large enough to eliminate finite-size
effects on the force.

The results, computed with BEM, are shown in~\figref{single-cylinder},
where all dimensions are given in units of $R$.  As our goal is to
examine the force relative to the $r = 0$ case, we normalize the force
results by the peak repulsive force along the $z$-axis, i.e.,
$F(r,z)/\mathrm{max}_z[F(r=0,z)]$.  Moving away from $r = 0$, we find
that this ratio is consistently larger than 1 for all $z$ plotted.
This increase can be quite large (up to 55 times larger) before the
force rapidly switches sign to attraction.  We propose the following
explanation for this effect: as $r \rightarrow R$, the fringing fields
of the thin plate, which are induced by dipolar charge/current
fluctuations of the cylinder, grow without bound.  This leads to an
enhanced plate-dipole coupling with increasing $r$, which for fixed
$z$ should lead to an enhanced repulsion if the force at that $z$ is
repulsive for $r = 0$.  However, when the cylinder-plate separation
becomes comparable to the cylinder height $h$, higher-order (i.e.,
non-dipolar) currents on the cylinder will begin to contribute to the
energy.  As these higher-order multipole currents are not bound by the
dipolar symmetry argument of~\citeasnoun{LevinMc10}, we expect their
contribution to the force to be attractive in general.  With this
interpretation, the force curves of~\figref{single-cylinder} show a
competition between the contribution of dipole currents on the
cylinder, which is repulsive for certain $z$ and grows without bound
as $r \rightarrow R$, and the contributions of higher-order cylinder
currents which are attractive and grow more rapidly than the dipole
term as $r \rightarrow R$.

This argument is supported by the results of~\figref{single-cylinder},
and raises a further interesting theoretical question: is it possible
to \emph{increase} the repulsive effect by \emph{decreasing} the size
of the cylinder (while keeping its aspect ratio constant)?  The
motivation for this is that a smaller cylinder can be brought closer
to the edge of the plate before higher order currents contribute
(attractively) to the force, therefore taking greater advantage of the
increased fringing fields and leading to a larger repulsive net force.
(Note that this does not violate PFA, because whenever we have
repulsion we are assuming $h$ is always comparable to the
surface-surface separation.)  Numerical computations (see below)
support this argument; however, a rigorous analysis is of course
required to confirm this.  As this is of course an unphysical limit,
we will not perform such an analysis here.  Rather, we simply use this
observation to illustrate that it is crucial to account for the finite
thickness $T$ of the plate: once $R-r \sim T$, a cylinder of any size
will induce higher-order currents on the plate, which will also lead
to attraction.  A non-zero value of $T$ must therefore be set in order
to obtain physically meaningful force bounds.

\begin{figure}[tb]
\includegraphics[width=1.0\columnwidth]{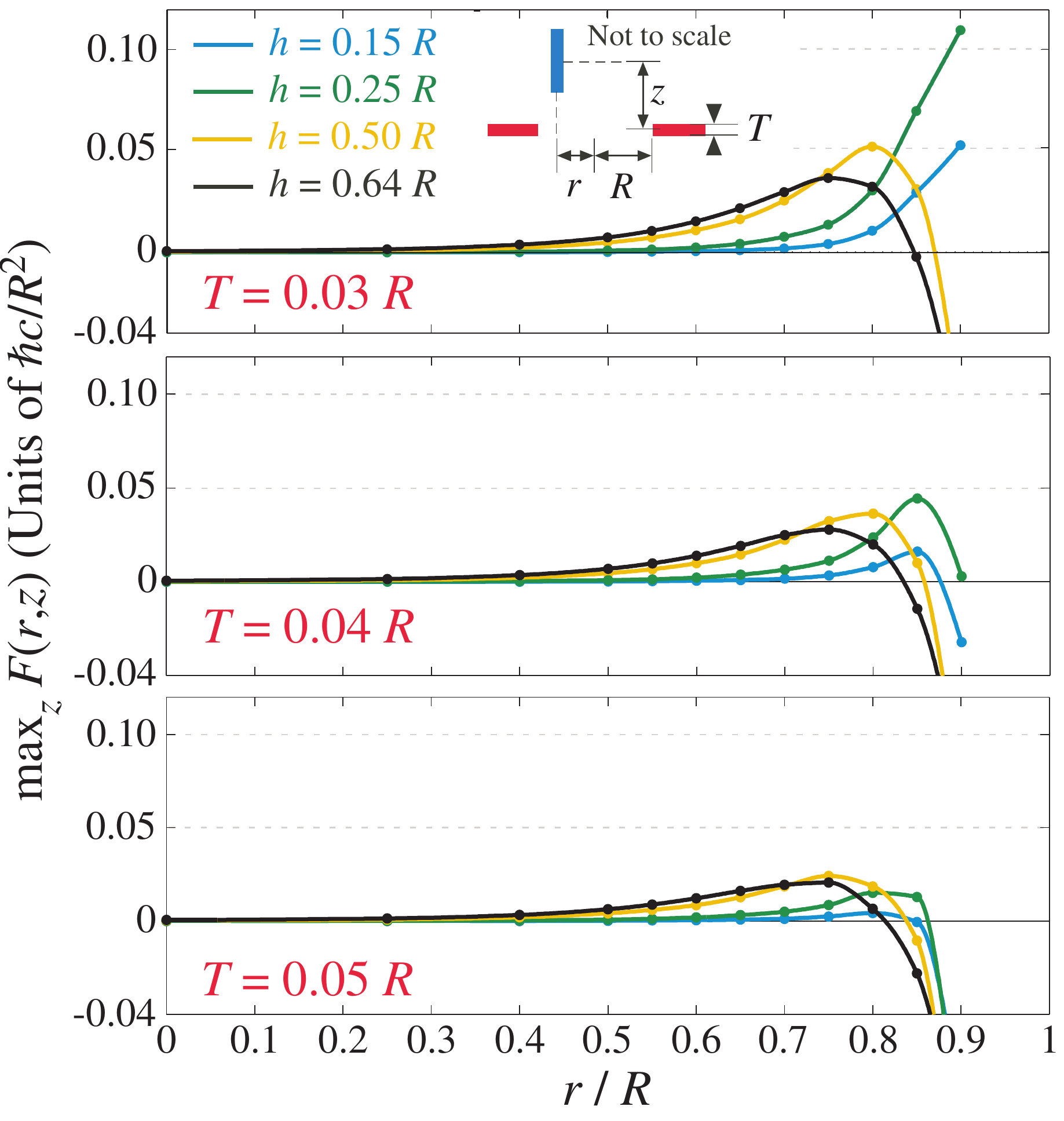}
\centering
\caption{The effect of plate thickness on the Casimir force on a
  single cylinder.  For each panel, the plate thickness is fixed and
  the Casimir force (maximized over $z>0$) on a single cylinder is
  plotted as a function of $x$.  The force for different values of
  cylinder height $h$ are shown (for each $h$, the aspect ratio of the
  cylinder is fixed).  For small $T$ (Top), decreasing the cylinder
  size can actually lead to an increased repulsion for $x\sim R$.
  Increasing $T$ (Middle and Bottom) leads to a decreased repulsion
  for all cylinders at all $x$, but the effect is strongest for small
  $h$ cylinders.}
\label{fig:thick-potential}
\end{figure}

We now consider the more realistic case of the force between a
cylinder an a finite thickness $(T>0)$ plate
in~\figref{thick-potential}.  In light of the argument presented
above, we also study the effect for several values of $h$.  On a
technical note, use of the boundary-element method requires that the
size of the mesh used to discretize the surfaces be smaller than any
characteristic separation.  However, now that $T > 0$, we must place
two copies of the annulus used above within a distance $T$ of each
other.  As $T \ll R$, this requires a much denser mesh than used above
in the thin plate case.  Roughly, the number of surface mesh elements
will scale as $T^{-2}$, so that the overall spatial and temporal
resource requirements scale as $\sim T^{-4}$ and $\sim T^{-6}$,
respectively~\cite{ReidRo09}.  Due to this rather steep scaling, we
find that for the range of $T$ considered here an annulus with outer
radius of $8 R$ proves too computationally demanding.  To reduce the
overall number of surface mesh elements, we truncate the outer radius
at $1.6 R$ instead; although there will now be finite-size effects, we
estimate their effect at well below $10\%$ of the total force in all
cases.  We examine three plate thicknesses $T/R = 0.03, 0.04, 0.05$
and five different values of $h/R$.  For each cylinder value of $h$,
the corresponding cylinder width is chosen to as to keep a constant
cylinder aspect ratio.  The results are shown
in~\figref{thick-potential}; to simplify the presentation, for each
$T$ we plot only the maximum repulsive force force $\max_{z>0}
F_z(r,z)$ for each $r$.  The top panel of~\figref{thick-potential}
supports the argument given in the previous paragraph: although for
most values of $r$, $h = 0.64 R$ gives the largest repulsive force, as
$r/R \rightarrow 1$, cylinders of progressively smaller size exhibit a
stronger repulsion.  (Although we expect that the limits $h\rightarrow
0$, $r/R \rightarrow 1$ and $T\rightarrow 0$ can be taken in such a
way as to make the peak repulsion unbounded, numerical limitations
make computation of forces with $r/R > 0.9$ difficult.)  On the other
hand, as $T$ increases the attractive contributions of higher-order
plate currents turn on at progressively smaller $r$.  Further, this
effect is strongest for the small-$h$ cylinders, as shown by
successive panels of~\figref{thick-potential}.  For example, by $T/R =
0.04$ the force curves for $h < 0.25 R$ now show attraction by $r =
0.9 R$, and for $T/R = 0.05$, the cylinders with $h/R \sim 0.5$ now
show the strongest repulsive effect when maximized over all $r$.  In
contrast to the strong $T$-dependence of the force for small $h$, for
$h = 0.64 R$ the results are quantitatively altered (approximately
$50\%$), but the order of magnitude of the repulsive effect remains.
On the basis of this, we conclude that for the range of thicknesses we
are interested in, the best value of $h/R$ is in the range $h/R \sim
0.5$.

\section{Multiple cylinders and uniaxial capsules}

\begin{figure}[tb]
\includegraphics[width=1.0\columnwidth]{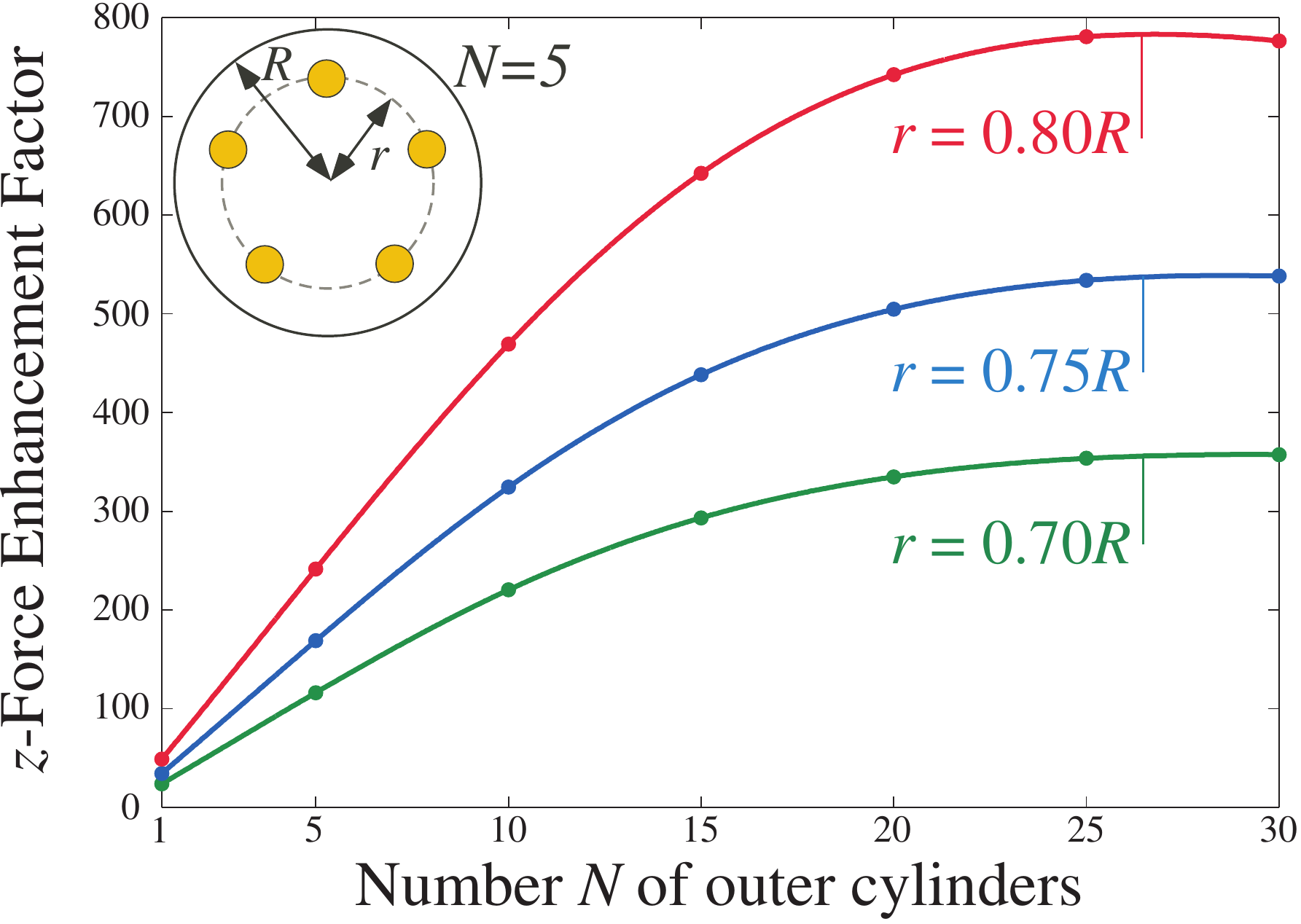}
\centering
\caption{Enhancement of Casimir repulsion with a a realistic geometry
  (\figref{configs}, Bottom Left): in place of a single cylinder, a
  ring of $N$ cylinders is placed at radius $r$ (shown for $N=5$ in
  the inset).  The cylinders have height $h/R = 0.64$, and for
  computational tractability the plate thickness $T$ is zero.  For
  small $N$ the force increases linearly, with slope corresponding
  approximately to the magnitude of enhancement
  of~\figref{single-cylinder}.  As $N$ increases, the effect
  saturates.  As in previous figures, the force is maximized over
  $z$.}
\label{fig:multiple-cylinders}
\end{figure}

Previously, we found that the repulsive force is greatly enhanced as a
cylinder is brought close to the edge of a plate---how close depended
on both the cylinder height $h$ and the plate thickness $T$.  It is
reasonable to assume that this effect should be enhanced if multiple
cylinders are now added, bearing aside non-additive cylinder-cylinder
interactions.  In particular, if $N$ cylinders are spread about a ring
of radius $r$, one expects that for $N$ not too large the repulsive
force should scale linearly with $N$.  As $N$ increases,
cylinder-cylinder interactions will take over.  In~\secref{multiples},
we will examine this behavior in detail.  We find that the
non-additive cylinder-cylinder interactions simply lead to a
saturation of the repulsive effect, and that the value of $N$ at which
this occurs is quite high, leading to a large force enhancement.  In
the limit of large $N$, it is appropriate to replace the ring of
cylinders with a homogeneous cylindrical capsule of uniaxial
conductivity~\figref{configs}(d).  This is examined
in~\secref{uniaxial}, and allows us to obtain an upper bound for the
repulsive force in our system.  In addition, as these systems have
rotational symmetry, the net force is along the $z$-axis and there is
no longer an attractive radial component.

\subsection{Multiple cylinders}
\label{sec:multiples}

We examine the effect of multiple cylinders with BEM simulations.  Due
to the computational limitations in BEM for a plate of finite
thickness, we cannot directly simulate multiple cylinders next to a
thick plate.  Instead, we utilize the fact
(see~\figref{thick-potential}) that a cylinder of $h = 0.64 R$ is not
strongly affected by a finite but small slab thickness and perform BEM
simulations for a thin plate instead.  We expect that these results
will still be approximately valid for plate thicknesses $T \lesssim
0.05 R$.  The results of these computations are shown
in~\figref{multiple-cylinders} for three values of ring radius $r/R$.
For small $N$, the $z$-force increases approximately linearly, with
the slope roughly corresponding to the size of the enhancement factor
for a single cylinder in~\figref{single-cylinder} (red curve) for the
value $x \sim r$.  As $N$ increases, non-additive cylinder-cylinder
interactions become important, and this linear enhancement eventually
saturates.  However, it is interesting that this saturation does not
occur until fairly large $N$, at which point the peak enhancement
factor is $\sim\,800$ for $r = 0.8 R$ (larger $r$ exhibit weaker
repulsion, in line with~\figref{single-cylinder}).  Further, it is
interesting to observe that increasing $N$ past this saturation point
does not have an appreciable effect on the force: the additional
cylinders are simply screened, and net repulsive force is practically
unchanged.  Additional calculations (not shown) for thinner cylinders
show very similar values for the peak repulsion, with saturation after
this peak.  This leads us to believe that the ideal uniaxial capsules,
shown in~\figref{configs}~(d), are actually a fairly good
approximation to the more realistic configuration of many cylinders
considered here.  The advantage of the uniaxial capsules, examined in
the next section, is that we can easily simulate them in FDTD and
incorporate the thick plate, which as stated before is crucial for
obtaining a realistic bound on the repulsion.

\subsection{Uniaxial capsules}
\label{sec:uniaxial}

The basic mechanism of the repulsive effect relies on combining
fringing fields with objects of high, anisotropic conductivity.  Until
this point, this anisotropy has been realized via the shape of the
objects.  However, as argued in the previous section, the limit in
which the cylinders are allowed to become arbitrarily narrow and
densely packed is well-defined and furthermore seems to be a good
approximation to the finite-sized cylinder configurations considered
previously.  In this limit, the cylindrical ring is replaced by a
cylindrical capsule of homogeneous material with uniaxial conductivity
along the $z$ axis: $\varepsilon(i\xi) = \mathrm{diag}(1,1,\infty)$.
The capsule has a height $h$, outer radius $r$, and (for generality)
inner radius $r^\prime$; see~\figref{configs}~(d).  Although we cannot
simulate anisotropic materials with our boundary-element method, they
can easily be treated with FDTD~\cite{RodriguezMc09:PRA,
  McCauleyRo10:PRA}.  This has the advantage that finite-thickness
plates can be incorporated with no additional computational cost, and
we can also exploit the cylindrical symmetry of the configuration to
make the problem two-dimensional.  In general the inner radius
$r^\prime$ can be non-zero.  However, we examined the difference in
the force on a capsule as the inner radius $r^\prime$ is varied, and
it turned out that this has almost no effect (less than $1\%$) on the
force - the inner conducting material is almost, if not entirely,
screened from the field of the plate.  Therefore, without loss of
generality in the following we set $r^\prime = 0$ and consider solid
cylinders only.

\begin{figure}[tb]
\includegraphics[width=1.0\columnwidth]{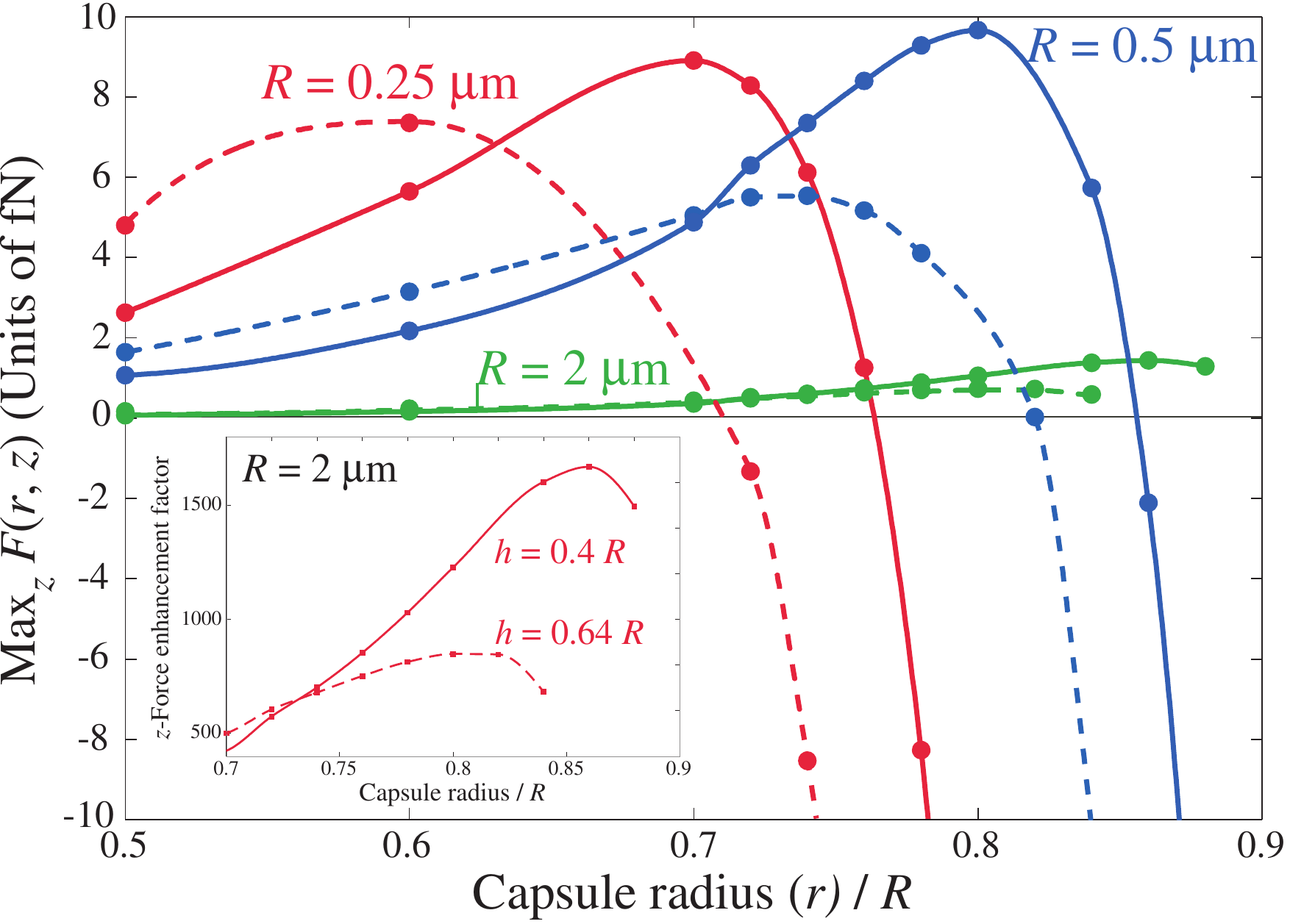}
\centering
\caption{Forces (maximized over $z$) for a thick cylindrical capsule
  composed of uniaxial conductor (\figref{configs}, Bottom Right).  In
  contrast to previous figures, here the plate thickness is fixed to
  $T = 20\,\mathrm{nm}$ and the hole radius $R$ is varied to three
  different values.  For each $R$, the force is computed, as a
  function of capsule radius $r$, for capsule heights $h = 0.5 R$
  (solid lines) and $h = 0.64 R$ (dashed lines).  \emph{Inset}:
  $z$-force enhancement factors as a function for both values of $h$
  for $R = 2\,\mu\mathrm{m}$, which approximates a thin plate.  The
  maximum force is approximately $10\,\mathrm{fN}$.}
\label{fig:thick-capsule}
\end{figure}

In the previous sections, we argued that the enhancement of Casimir
repulsion in between the point dipole and PFA regimes is due to two
effects - the increased size of the center object, and the increase in
the fringing fields of the plate.  We did this by separately examining
each effect in turn.  We are now in a position to examine the full
combination of these effects, primarily to obtain an approximate upper
bound on the possible strength of the repulsive effect assuming a
fixed plate thickness $T$.  To preserve contact with our previous
work~\citeasnoun{LevinMc10}, in this section we fix our units so that
$T = 20\,\mathrm{nm}$ in all cases.  From the results of previous
sections, maximizing the repulsion over the remaining parameters (hole
radius $R$, capsule radius $r$, capsule height $h$--here it is
understood that we have already maximized over vertical displacement
$z>0$) will yield a finite peak repulsion.  To simply the
presentation, we will write both $h$ and $r$ in units of the hole
width $R$, so that only the radios $h/R$ and $r/R$ are relevant.  Then
for fixed $h/R$ and $r/R$, as $R$ is varied (keeping
$T=20\,\mathrm{nm}$ always) we expect the force to scale as $R^{-2}$
for $R \gg T$; as $R$ decreases further, we expect the effect of $T>0$
to overwhelm the $R^{-2}$ scaling at some point, leaving an optimal
value of $R$ for each $h/R$ and $r/R$.  From the results of the
previous section, we expect that examining both $h/R = 0.4$ and $h/R =
0.64$ will give a good estimate for the maximum force enhancement.
For these two values, we consider three values of $R$, such that $T/R
= 0.0025, 0.01$, and 0.02 (corresponding to $R = 2\,\mu\mathrm{m},
0.5\,\mu\mathrm{m}, 0.25\,\mu\mathrm{m}$, respectively).
In~\figref{thick-capsule} we plot the force for these six
configurations as a function of capsule radius $r$.  The case $R =
2\,\mu\mathrm{m}$ is nearly indistinguishable from the thin plate
case.  To make contact with the results presented in previous
sections, we plot the enhancement factor in the force for this case in
the inset of~\figref{thick-capsule}.  For $h/R=0.64$, we get an
enhancement similar to the previous section with multiple cylinders,
while for $h/R = 0.4$ this factor is $\sim 1600$.  Of course we know
from previous results that this value can be made arbitrarily large;
if we are to fix the plate thickness, we should consider instead the
absolute magnitude of the repulsive force.  When this is done, we see
that going from $R=2\,\mu\mathrm{m}$ to $R=0.5\,\mu\mathrm{m}$ yields
a further enhancement factor of 6, whereas for the $T=0$ case it
should be 16.  When $R$ is further decreased we see a decrease in the
force for $h/R = 0.5$, and only a slight increase for $h/R=0.64$.
Further decreases in $R$ (not shown) lead to a decrease in the force
for both $h$.  We see that although the enhancement factor for $R =
2\,\mu\mathrm{m}$ is much larger for $h/R = 0.4$ than $h/R=0.64$, the
actual peak magnitudes for the forces are fairly similar for both
values of $h$, and peak at approximately $10\,\mathrm{fN}$.  This is
to be contrasted with the peak repulsive force found
in~\citeasnoun{LevinMc10}, which was approximately $10\,\mathrm{aN}$,
and gives our quoted enhancement factor of $\sim 10^3$.

Additionally, we can use the uniaxial capsule configuration to examine
the importance of our assumption of perfect conductors.  Replacing the
conducting plate with gold (plasma frequency $1.37\times
10^{16}\,\mathrm{rad/s}$) and the capsule with a material that has
permittivity $\varepsilon(\omega) =
\mathrm{diag}(1,1,\varepsilon_\mathrm{gold}(\omega))$, we find that
the peak repulsive force for $R = 2\,\mu\mathrm{m}$ is reduced by
$50\%$, but that for $R = 500\,\mathrm{nm}$ it is only reduced by
$20\%$, while for $R = 250\,\mathrm{nm}$ it is \emph{increased} by
$10\%$.  The increase in the repulsive force for smaller $R$ can be
understood from the fact (mentioned in~\cite{LevinMc10}) that the
attractive forces from finite-thickness plates come from high-$\omega$
components, which are cutoff by the plasma frequency in the gold
permittivity.  Therefore, using gold actually helps to attenuate some
of the attractive effects introduced by using finite-thickness plates,
and implies that our force bounds remain valid for imperfect
conductors.  Finally, we examined the effect of introducing a long
dielectric column attached to the capsule (this would serve as both an
anchor for the capsule and a means of detecting the force on it).  We
find that a column made of silica (modeled as constant permittivity
$\varepsilon = 2.25$) and of half the width of the capsule does not
significantly modify the repulsive force.  However, taking the column
width equal to the capsule width reduces the peak repulsive force by a
factor of 10.

\section{Conclusions}

We have shown that Casimir repulsion is not limited to the
small-particle dipole-interaction regime of our previous paper, and
can in fact be considerably enhanced by considering interactions
between macroscopic materials with highly anisotropic microstructures.
As a practical matter, severe challenges remain in experimental
realization of Casimir repulsion via this particular geometry.  For an
individual capsule with realistic conductivity, the force is only
$10\,\mathrm{fN}$, which, while 1000 times larger than the result of
our previous paper, is still below the detection threshold of atomic
force microscopy~\cite{Goddard07}.  If this hole/capsule geometry were
arranged in a periodic array with period twice the hole diameter (our
numerical calculations indicate that this is sufficient to prevent
interactions across holes), the repulsive pressure would be
approximately $2\,\mathrm{mPa}$ for $R = 500\,\mu\mathrm{m}$,
comparable to the attractive force between metal plates at
$900\,\mathrm{nm}$ separation, but the fabrication and alignment of
such a structure appears extremely difficult.  As a more general
point, however, we believe that one route to obtaining a number of
exotic Casimir effects would be to exploit effective uniaxial
conductors formed by, e.g., vertical arrays of
nanowires~\cite{Thelander06}.  Even if the repulsive regime cannot be
achieved, our results suggest that significant modifications to the
Casimir force may result.  Furthermore, other work has shown that
additional exotic effects can be achieved by anisotropic patterning,
such as orientation induced force transitions and
torques~\cite{McCauley11:orientation}, and anisotropic effects in many
other geometries remain to be explored both theoretically and
experimentally.

Near the completion of this work, a related
paper~\cite{Milton11:repulsion} appeared which also examines (via
analytical calculations) Casimir repulsion in geometries beyond the
thin-plate/dipole regime, finding for example that repulsion persists
when the plate has a wedge-like perpendicular profile as opposed to a
thin line segment as considered here.


\begin{thebibliography}{26}
\expandafter\ifx\csname natexlab\endcsname\relax\def\natexlab#1{#1}\fi
\expandafter\ifx\csname bibnamefont\endcsname\relax
  \def\bibnamefont#1{#1}\fi
\expandafter\ifx\csname bibfnamefont\endcsname\relax
  \def\bibfnamefont#1{#1}\fi
\expandafter\ifx\csname citenamefont\endcsname\relax
  \def\citenamefont#1{#1}\fi
\expandafter\ifx\csname url\endcsname\relax
  \def\url#1{\texttt{#1}}\fi
\expandafter\ifx\csname urlprefix\endcsname\relax\def\urlprefix{URL }\fi
\providecommand{\bibinfo}[2]{#2}
\providecommand{\eprint}[2][]{\url{#2}}

\bibitem[{\citenamefont{Levin et~al.}(2010)\citenamefont{Levin, McCauley,
  Rodriguez, Reid, and Johnson}}]{LevinMc10}
\bibinfo{author}{\bibfnamefont{M.}~\bibnamefont{Levin}},
  \bibinfo{author}{\bibfnamefont{A.~P.} \bibnamefont{McCauley}},
  \bibinfo{author}{\bibfnamefont{A.~W.} \bibnamefont{Rodriguez}},
  \bibinfo{author}{\bibfnamefont{M.~T.~Homer} \bibnamefont{Reid}},
  \bibnamefont{and} \bibinfo{author}{\bibfnamefont{S.~G.}
  \bibnamefont{Johnson}}, \bibinfo{journal}{Phys. Rev. Lett.}
  \textbf{\bibinfo{volume}{105}}, \bibinfo{pages}{090403}
  (\bibinfo{year}{2010}).

\bibitem[{\citenamefont{Milton}(2001)}]{milton01}
\bibinfo{author}{\bibfnamefont{K.~A.} \bibnamefont{Milton}},
  \emph{\bibinfo{title}{The {Casimir} Effect: Physical Manifestations of
  Zero-Point Energy}} (\bibinfo{publisher}{Singapore: World Scientific},
  \bibinfo{year}{2001}).

\bibitem[{\citenamefont{Rodriguez et~al.}(2011)\citenamefont{Rodriguez,
  Capasso, and Johnson}}]{RodriguezCa11}
\bibinfo{author}{\bibfnamefont{A.~W.} \bibnamefont{Rodriguez}},
  \bibinfo{author}{\bibfnamefont{F.}~\bibnamefont{Capasso}}, \bibnamefont{and}
  \bibinfo{author}{\bibfnamefont{S.~G.} \bibnamefont{Johnson}},
  \bibinfo{journal}{Nature Phot.} \textbf{\bibinfo{volume}{5}},
  \bibinfo{pages}{211} (\bibinfo{year}{2011}).

\bibitem[{\citenamefont{Derjaguin}(1934)}]{Derjaguin34}
\bibinfo{author}{\bibfnamefont{B.~V.} \bibnamefont{Derjaguin}},
  \bibinfo{journal}{Kolloid Z.} \textbf{\bibinfo{volume}{69}},
  \bibinfo{pages}{155} (\bibinfo{year}{1934}).

\bibitem[{\citenamefont{van Enk}(1995)}]{Enk95:torque}
\bibinfo{author}{\bibfnamefont{S.~J.} \bibnamefont{van Enk}},
  \bibinfo{journal}{Phys. Rev.~A} \textbf{\bibinfo{volume}{52}},
  \bibinfo{pages}{2569} (\bibinfo{year}{1995}).

\bibitem[{\citenamefont{Shao et~al.}(2005)\citenamefont{Shao, Tong, and
  Luo}}]{Shao05}
\bibinfo{author}{\bibfnamefont{C.-G.} \bibnamefont{Shao}},
  \bibinfo{author}{\bibfnamefont{A.-H.} \bibnamefont{Tong}}, \bibnamefont{and}
  \bibinfo{author}{\bibfnamefont{J.}~\bibnamefont{Luo}},
  \bibinfo{journal}{Phys. Rev.~A} \textbf{\bibinfo{volume}{72}},
  \bibinfo{pages}{022102} (\bibinfo{year}{2005}).

\bibitem[{\citenamefont{Munday et~al.}(2005)\citenamefont{Munday, Iannuzzi,
  Barash, and Capasso}}]{Munday05}
\bibinfo{author}{\bibfnamefont{J.~N.} \bibnamefont{Munday}},
  \bibinfo{author}{\bibfnamefont{D.}~\bibnamefont{Iannuzzi}},
  \bibinfo{author}{\bibfnamefont{Y.}~\bibnamefont{Barash}}, \bibnamefont{and}
  \bibinfo{author}{\bibfnamefont{F.}~\bibnamefont{Capasso}},
  \bibinfo{journal}{Phys. Rev.~A} \textbf{\bibinfo{volume}{71}},
  \bibinfo{pages}{042102} (\bibinfo{year}{2005}).

\bibitem[{\citenamefont{Kenneth and Nussinov}(2001)}]{Kenneth00:torque}
\bibinfo{author}{\bibfnamefont{O.}~\bibnamefont{Kenneth}} \bibnamefont{and}
  \bibinfo{author}{\bibfnamefont{S.}~\bibnamefont{Nussinov}},
  \bibinfo{journal}{Phys. Rev.~D} \textbf{\bibinfo{volume}{63}},
  \bibinfo{pages}{121701({R})} (\bibinfo{year}{2001}).

\bibitem[{\citenamefont{Rodrigues et~al.}(2006)\citenamefont{Rodrigues,
  Maia~Neto, Lambrecht, and Reynaud}}]{Rodrigues06:torque}
\bibinfo{author}{\bibfnamefont{R.~B.} \bibnamefont{Rodrigues}},
  \bibinfo{author}{\bibfnamefont{P.~A.} \bibnamefont{Maia~Neto}},
  \bibinfo{author}{\bibfnamefont{A.}~\bibnamefont{Lambrecht}},
  \bibnamefont{and} \bibinfo{author}{\bibfnamefont{S.}~\bibnamefont{Reynaud}},
  \bibinfo{journal}{Europhys. Lett.} \textbf{\bibinfo{volume}{75}},
  \bibinfo{pages}{822} (\bibinfo{year}{2006}).

\bibitem[{\citenamefont{McCauley et~al.}(2011)\citenamefont{McCauley, Rosa,
  Rodriguez, Joannopoulos, Dalvit, and Johnson}}]{McCauley11:orientation}
\bibinfo{author}{\bibfnamefont{A.~P.} \bibnamefont{McCauley}},
  \bibinfo{author}{\bibfnamefont{F.~S.~S.} \bibnamefont{Rosa}},
  \bibinfo{author}{\bibfnamefont{A.~W.} \bibnamefont{Rodriguez}},
  \bibinfo{author}{\bibfnamefont{J.~D.} \bibnamefont{Joannopoulos}},
  \bibinfo{author}{\bibfnamefont{D.~A.~R.} \bibnamefont{Dalvit}},
  \bibnamefont{and} \bibinfo{author}{\bibfnamefont{S.~G.}
  \bibnamefont{Johnson}}, \bibinfo{journal}{Phys. Rev. A, \emph{In Press}}
  (\bibinfo{year}{2011}).

\bibitem[{\citenamefont{Emig et~al.}(2003)\citenamefont{Emig, Hanke,
  Golestanian, and Kardar}}]{emig03_1}
\bibinfo{author}{\bibfnamefont{T.}~\bibnamefont{Emig}},
  \bibinfo{author}{\bibfnamefont{A.}~\bibnamefont{Hanke}},
  \bibinfo{author}{\bibfnamefont{R.}~\bibnamefont{Golestanian}},
  \bibnamefont{and} \bibinfo{author}{\bibfnamefont{M.}~\bibnamefont{Kardar}},
  \bibinfo{journal}{Phys. Rev.~A} \textbf{\bibinfo{volume}{67}},
  \bibinfo{pages}{022114} (\bibinfo{year}{2003}).

\bibitem[{\citenamefont{Emig}(2007)}]{Emig07:ratchet}
\bibinfo{author}{\bibfnamefont{T.}~\bibnamefont{Emig}}, \bibinfo{journal}{Phys.
  Rev. Lett.} \textbf{\bibinfo{volume}{98}}, \bibinfo{pages}{160801}
  (\bibinfo{year}{2007}).

\bibitem[{\citenamefont{Maia~Neto et~al.}(2008)\citenamefont{Maia~Neto,
  Lambrecht, and Reynaud}}]{Neto08}
\bibinfo{author}{\bibfnamefont{P.~A.} \bibnamefont{Maia~Neto}},
  \bibinfo{author}{\bibfnamefont{A.}~\bibnamefont{Lambrecht}},
  \bibnamefont{and} \bibinfo{author}{\bibfnamefont{S.}~\bibnamefont{Reynaud}},
  \bibinfo{journal}{Phys. Rev.~A} \textbf{\bibinfo{volume}{78}},
  \bibinfo{pages}{012115} (\bibinfo{year}{2008}).

\bibitem[{\citenamefont{Lambrecht and Marachevsky}(2008)}]{Lambrecht09}
\bibinfo{author}{\bibfnamefont{A.}~\bibnamefont{Lambrecht}} \bibnamefont{and}
  \bibinfo{author}{\bibfnamefont{V.~N.} \bibnamefont{Marachevsky}},
  \bibinfo{journal}{Phys. Rev. Lett.} \textbf{\bibinfo{volume}{101}},
  \bibinfo{pages}{160403} (\bibinfo{year}{2008}).

\bibitem[{\citenamefont{Chiu et~al.}(2010)\citenamefont{Chiu, Klimchitskaya,
  Marachevsky, Mostepanenko, and Mohideen}}]{Chiu10}
\bibinfo{author}{\bibfnamefont{H.-C.} \bibnamefont{Chiu}},
  \bibinfo{author}{\bibfnamefont{G.~L.} \bibnamefont{Klimchitskaya}},
  \bibinfo{author}{\bibfnamefont{V.~N.} \bibnamefont{Marachevsky}},
  \bibinfo{author}{\bibfnamefont{V.~M.} \bibnamefont{Mostepanenko}},
  \bibnamefont{and} \bibinfo{author}{\bibfnamefont{U.}~\bibnamefont{Mohideen}},
  \bibinfo{journal}{Phys. Rev. B} \textbf{\bibinfo{volume}{81}},
  \bibinfo{pages}{115417} (\bibinfo{year}{2010}).

\bibitem[{\citenamefont{Leonhardt and Philbin}(2007)}]{Leonhardt07}
\bibinfo{author}{\bibfnamefont{U.}~\bibnamefont{Leonhardt}} \bibnamefont{and}
  \bibinfo{author}{\bibfnamefont{T.~G.} \bibnamefont{Philbin}},
  \bibinfo{journal}{New J. Phys.} \textbf{\bibinfo{volume}{9}},
  \bibinfo{pages}{254} (\bibinfo{year}{2007}).

\bibitem[{\citenamefont{Romanowsky and Capasso}(2008)}]{Romanowsky08}
\bibinfo{author}{\bibfnamefont{M.~B.} \bibnamefont{Romanowsky}}
  \bibnamefont{and} \bibinfo{author}{\bibfnamefont{F.}~\bibnamefont{Capasso}},
  \bibinfo{journal}{Phys. Rev.~A} \textbf{\bibinfo{volume}{78}},
  \bibinfo{pages}{042110} (\bibinfo{year}{2008}).

\bibitem[{\citenamefont{Rosa}(2009)}]{Rosa09}
\bibinfo{author}{\bibfnamefont{F.~S.~S.} \bibnamefont{Rosa}},
  \bibinfo{journal}{J. Phys. Conf. Ser.} \textbf{\bibinfo{volume}{161}},
  \bibinfo{pages}{012039} (\bibinfo{year}{2009}).

\bibitem[{\citenamefont{Zhao et~al.}(2009)\citenamefont{Zhao, Zhou, Koschny,
  Economou, and Soukoulis}}]{Zhao09}
\bibinfo{author}{\bibfnamefont{R.}~\bibnamefont{Zhao}},
  \bibinfo{author}{\bibfnamefont{J.}~\bibnamefont{Zhou}},
  \bibinfo{author}{\bibfnamefont{T.}~\bibnamefont{Koschny}},
  \bibinfo{author}{\bibfnamefont{E.~N.} \bibnamefont{Economou}},
  \bibnamefont{and} \bibinfo{author}{\bibfnamefont{C.~M.}
  \bibnamefont{Soukoulis}}, \bibinfo{journal}{Phys. Rev. Lett.}
  \textbf{\bibinfo{volume}{103}}, \bibinfo{pages}{103602}
  (\bibinfo{year}{2009}).

\bibitem[{\citenamefont{Reid et~al.}(2009)\citenamefont{Reid, Rodriguez, White,
  and Johnson}}]{ReidRo09}
\bibinfo{author}{\bibfnamefont{M.~T.~Homer} \bibnamefont{Reid}},
  \bibinfo{author}{\bibfnamefont{A.~W.} \bibnamefont{Rodriguez}},
  \bibinfo{author}{\bibfnamefont{J.}~\bibnamefont{White}}, \bibnamefont{and}
  \bibinfo{author}{\bibfnamefont{S.~G.} \bibnamefont{Johnson}},
  \bibinfo{journal}{Phys. Rev. Lett.} \textbf{\bibinfo{volume}{103}},
  \bibinfo{pages}{040401} (\bibinfo{year}{2009}).

\bibitem[{\citenamefont{Rodriguez et~al.}(2009)\citenamefont{Rodriguez,
  McCauley, Joannopoulos, and Johnson}}]{RodriguezMc09:PRA}
\bibinfo{author}{\bibfnamefont{A.~W.} \bibnamefont{Rodriguez}},
  \bibinfo{author}{\bibfnamefont{A.~P.} \bibnamefont{McCauley}},
  \bibinfo{author}{\bibfnamefont{J.~D.} \bibnamefont{Joannopoulos}},
  \bibnamefont{and} \bibinfo{author}{\bibfnamefont{S.~G.}
  \bibnamefont{Johnson}}, \bibinfo{journal}{Phys. Rev.~A}
  \textbf{\bibinfo{volume}{80}}, \bibinfo{pages}{012115}
  (\bibinfo{year}{2009}).

\bibitem[{\citenamefont{McCauley et~al.}(2010)\citenamefont{McCauley,
  Rodriguez, Joannopoulos, and Johnson}}]{McCauleyRo10:PRA}
\bibinfo{author}{\bibfnamefont{A.~P.} \bibnamefont{McCauley}},
  \bibinfo{author}{\bibfnamefont{A.~W.} \bibnamefont{Rodriguez}},
  \bibinfo{author}{\bibfnamefont{J.~D.} \bibnamefont{Joannopoulos}},
  \bibnamefont{and} \bibinfo{author}{\bibfnamefont{S.~G.}
  \bibnamefont{Johnson}}, \bibinfo{journal}{Phys. Rev.~A}
  \textbf{\bibinfo{volume}{81}}, \bibinfo{pages}{012119}
  (\bibinfo{year}{2010}).

\bibitem[{\citenamefont{Rahi et~al.}(2010)\citenamefont{Rahi, Kardar, and
  Emig}}]{Rahi10:PRL}
\bibinfo{author}{\bibfnamefont{S.~J.} \bibnamefont{Rahi}},
  \bibinfo{author}{\bibfnamefont{M.}~\bibnamefont{Kardar}}, \bibnamefont{and}
  \bibinfo{author}{\bibfnamefont{T.}~\bibnamefont{Emig}},
  \bibinfo{journal}{Phys. Rev. Lett.} \textbf{\bibinfo{volume}{105}},
  \bibinfo{pages}{070404} (\bibinfo{year}{2010}).

\bibitem[{\citenamefont{Goddard et~al.}(2007)\citenamefont{Goddard, Brenner,
  Lyshevski, and Iafrate}}]{Goddard07}
\bibinfo{editor}{\bibfnamefont{W.~A.} \bibnamefont{Goddard}},
  \bibinfo{editor}{\bibfnamefont{D.}~\bibnamefont{Brenner}},
  \bibinfo{editor}{\bibfnamefont{S.~E.} \bibnamefont{Lyshevski}},
  \bibnamefont{and} \bibinfo{editor}{\bibfnamefont{G.~J.}
  \bibnamefont{Iafrate}}, eds. (\bibinfo{publisher}{CRC Press},
  \bibinfo{address}{Boca Raton, FL}, \bibinfo{year}{2007}),
  \bibinfo{edition}{2nd} ed.

\bibitem[{\citenamefont{Thelander et~al.}(2006)\citenamefont{Thelander,
  Agarwal, Brongersma, Eymery, Feiner, Forchel, Scheffler, Reiss, Ohlsson,
  Goesele et~al.}}]{Thelander06}
\bibinfo{author}{\bibfnamefont{C.}~\bibnamefont{Thelander}},
  \bibinfo{author}{\bibfnamefont{P.}~\bibnamefont{Agarwal}},
  \bibinfo{author}{\bibfnamefont{S.}~\bibnamefont{Brongersma}},
  \bibinfo{author}{\bibfnamefont{J.}~\bibnamefont{Eymery}},
  \bibinfo{author}{\bibfnamefont{L.~F.} \bibnamefont{Feiner}},
  \bibinfo{author}{\bibfnamefont{A.}~\bibnamefont{Forchel}},
  \bibinfo{author}{\bibfnamefont{M.}~\bibnamefont{Scheffler}},
  \bibinfo{author}{\bibfnamefont{W.}~\bibnamefont{Reiss}},
  \bibinfo{author}{\bibfnamefont{B.~J.} \bibnamefont{Ohlsson}},
  \bibinfo{author}{\bibfnamefont{U.}~\bibnamefont{Goesele}},
  \bibnamefont{et~al.}, \bibinfo{journal}{Mat. Today}
  \textbf{\bibinfo{volume}{9}}, \bibinfo{pages}{28} (\bibinfo{year}{2006}).

\bibitem[{\citenamefont{Milton et~al.}(2011)\citenamefont{Milton, Abalo,
  Parashar, Pourtolami, Brevik, and Ellingsen}}]{Milton11:repulsion}
\bibinfo{author}{\bibfnamefont{K.~A.} \bibnamefont{Milton}},
  \bibinfo{author}{\bibfnamefont{E.~K.} \bibnamefont{Abalo}},
  \bibinfo{author}{\bibfnamefont{P.}~\bibnamefont{Parashar}},
  \bibinfo{author}{\bibfnamefont{N.}~\bibnamefont{Pourtolami}},
  \bibinfo{author}{\bibfnamefont{I.}~\bibnamefont{Brevik}}, \bibnamefont{and}
  \bibinfo{author}{\bibfnamefont{S.~A.} \bibnamefont{Ellingsen}},
  \bibinfo{journal}{arXiv:hep-th/1103.4386}  (\bibinfo{year}{2011}).

\end{thebibliography}
\end{document}